\documentclass{article}  
\usepackage{abstract_2e} 

\usepackage[version=3]{mhchem}
\newcommand{\subhead}[1]{\vspace{0.2cm}\noindent{\emph{#1}}}

\usepackage{times}
\usepackage{epsfig}

\begin{document}  


\title{The Inhabitance Paradox: how habitability and inhabitancy are inseparable}


\author{Colin Goldblatt}
\affil{University of Victoria, Victoria, B.C., Canada (czg@uvic.ca)}


\runningtitle{The Inhabitance Paradox}

\titlemake  

\begin{abstracttext}
\section*{Introduction}
The dominant paradigm in assigning ``habitability'' to terrestrial planets is to define a circumstellar habitable zone: the locus of orbital radii in which the planet is neither too hot nor too cold for life as we know it. One dimensional climate models have identified theoretically impressive boundaries for this zone: a runaway greenhouse or water loss at the inner edge (Venus), and low-latitude glaciation followed by formation of \ce{CO2} clouds at the outer edge. A cottage industry now exists to ``refine'' the definition of these boundaries each year to the third decimal place of an AU. Using the same class of climate model, I show that the different climate states can overlap very substantially and that ``snowball Earth'', moist temperate climate, hot moist climate and a post-runaway dry climate can all be stable under the same solar flux. The radial extent of the temperate climate band is very narrow for pure water atmospheres, but can be widened with di-nitrogen and carbon dioxide. The width of the habitable zone is thus determined by the atmospheric inventories of these gases. Yet Earth teaches us that these abundances are very heavily influenced (perhaps even controlled) by biology. This is paradoxical: the habitable zone seeks to define the region a planet should be capable of harbouring life; yet whether the planet is inhabited will determine whether the climate may be habitable at any given distance from the star. This matters, because future life detection missions may use habitable zone boundaries in mission design. 

\section*{The habitable zone and physical climate}

A typical description of the ``habitable zone'' is the region of circumstellar space in which a planet might be habitable. Rather by convention, the practical definition has become the existence of liquid water, for life as we know it (that is: Earth) relies on this. 

As our conceptualization relies on the existence of on a certain phase of a certain chemical, our problem becomes finding the conditions of the physical climate of a planet which would put the surface and atmosphere in a desirable region of the pressure--temperature space of the phase diagram. Necessarily, we need to consider all three phases (all exist on Earth). This becomes a rather rich problem, for the different phases interact differently with electromagnetic radiation. There are, therefore, various climate feedbacks which express through water. 

The framework of dynamical systems theory becomes fundamental, as climate state depends on history; our task is to look for stable steady states of climate. To make progress, we can separately consider the fate of solar photons incident on Earth which may transfer energy to us, and the escape of thermally emitted photons which transfer energy away. Where energy fluxes are equal, there exists a steady state. Steady states which are stable to a small perturbation are habitable zone candidates. Thus, in Figures 1 and 2, we derive bifurcation diagrams for Earth's climate, mapping climate states as a function of circumstellar distance. There exist ice-covered, temperate moist, hot moist and hot dry climate states. The moist states both enjoy liquid water surfaces, but the sea-level temperature conditions are only favourable for life-as-we-know-it in the temperate moist state.

\begin{figure}
\begin{center}
\includegraphics[width=\columnwidth]{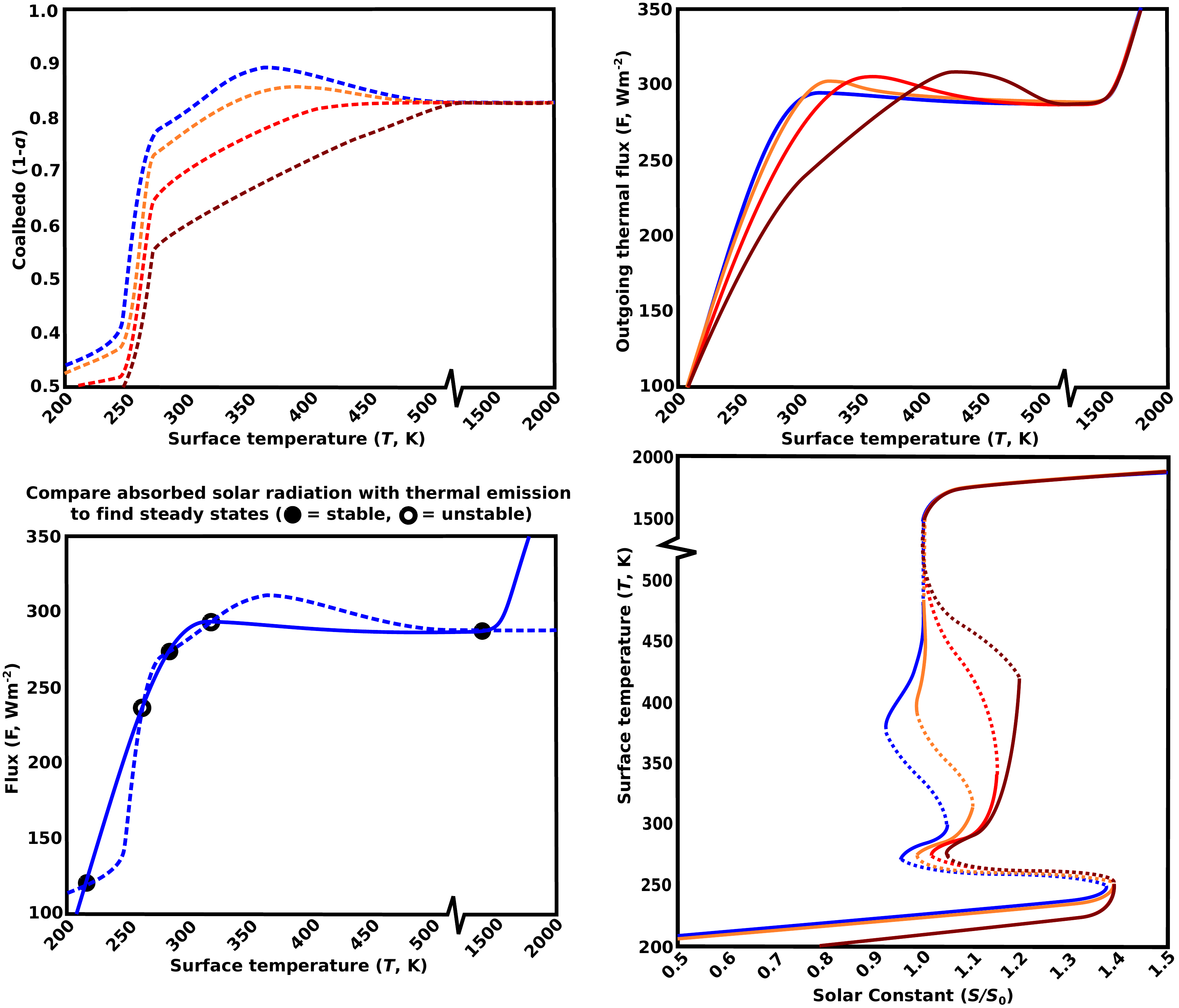}
\end{center}
\caption{Graphical derivation of steady states of climate with varying background gas inventories: none / pure water (blue) with orange through brown representing increasing p\ce{N2}. (a) Planetary co-albedo (b) Outgoing thermal flux (c) Example case of finding steady states by balancing outgoing thermal flux with the product of incident solar flux and co-albedo (d) Consequent bifurcation diagram, with steady states as a function of distance from the Sun, stable in solid lines and unstable dotted. These figures are schematics, sketched from my previous numerical results\cite{Goldblatt2013, Goldblatt2015}}
\end{figure}

\begin{figure}
\begin{center}
\includegraphics[width=\columnwidth]{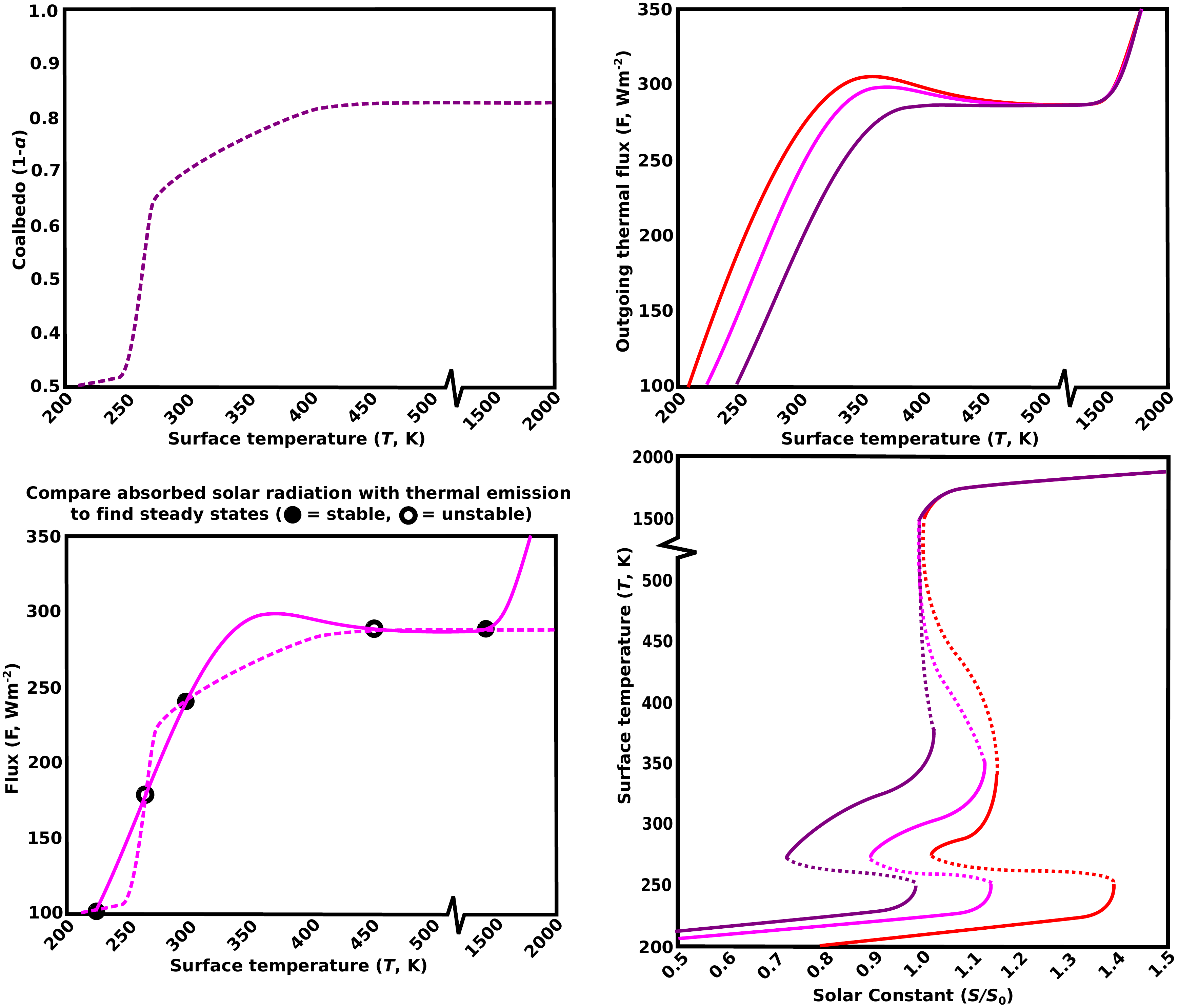}
\end{center}
\caption{Graphical derivation of steady states of climate with varying greenhouse gas inventories: red with some \ce{N2} but no \ce{CO2}, pink through purple representing increasing p\ce{CO2}. Other description as Figure 1.}
\end{figure}

For the fundamentals of how energy fluxes change as a response to water inventory (controlled by surface temperature via the Clausius-Claperon equation), I defer to descriptions in my previous papers\cite{Goldblatt2012a,Goldblatt2013,Goldblatt2015}. Herein, the focus is on how variation in atmospheric composition modifies these. 

Gases may be separated into \emph{radiatively active} and \emph{background} gases. The former absorb photons directly, and the most important subset of these, \emph{greenhouse gases}, does so in the infrared; carbon dioxide is the prototype for Earth. The latter absorb little radiation directly but will broaden the absorption of the radiative gases and scatter solar photons; di-nitrogen is the prototype for Earth. 

More background gas means less energy absorbed from the Sun because of higher albedo, and less or more thermal energy emitted depending on temperature (less at low temperatures where pressure-broadening is dominant, more at intermediate temperature where dilution of water aloft lowers the emission level). Consequent is the wet temperate climate state moving to higher solar constants and broadening: the habitable zone is wider and further from the Sun. 

More greenhouse gas simply reduces outgoing thermal radiation, and only does so significantly at lower temperatures. Thus the wet temperate climate state is moved to lower solar constants and is broader: the habitable zone is wider and nearer to the Sun.

\section*{Controls on atmospheric composition}

\vspace{-0.2cm}
\subhead{Is there an atmosphere?}

There are two kinds of planets, those with atmospheres and those without. The planets with atmospheres are simply those which have not lost them. This can be seen clearly on the ``Zahnle diagram'': with axis of escape velocity (planet mass) and solar heating, a simple power law empirically separates the two classes\cite{Zahnle2008,Zahnle2013}. 

\subhead{Carbon Dioxide}

There are two end member cases of how a planet may store a carbon dioxide reservoir, represented in our solar system by Earth and Venus. On Earth, there is a small atmospheric reservoir and a larger ocean reservoir, but the vast majority of oxidized carbon is stored in carbonate rocks. On Venus, the atmospheric carbon dioxide inventory is roughly equivalent to the contents of Earth's carbonate rocks, but there is neither an ocean inventory (as there is no ocean) nor any evidence of carbonate rocks. 

Let us consider how oxidized carbon is partitioned between reservoirs on an Earth-like planet. The partial pressure of atmospheric \ce{CO2} and the dissolved concentration in the ocean are in direct proportion, described by Henry's Law. But dissolved \ce{CO2} is not the major reservoir; in aqueous solution, the dissolution products of carbonic acid, bicarbonate ion and carbonate ion, will often be larger reservoirs. Conservation of mass is described via Dissolved Inorganic Carbon
(\ce{DIC = [CO2] + [HCO3^-] + [CO3^{2-}]}). Conservation of change of weak acids occurs through alkalinity, which balances net positive change from salts
(\ce{Alk = [HCO3^-] + 2[CO3^{2-}] + \ldots =  [Na^+] -[ Cl^-] + [Mg^2+] + [Ca^2+] + \ldots}). For some DIC, alkalinity will control the partitioning between different reservoirs; low alkalinity requires high \ce{[CO2]} and low \ce{[CO3^{2-}]}, high alkalinity the opposite. At intermediate alkalinity \ce{[HCO3^{-}]} dominates. Thus, for given DIC, alkalinity determines the atmospheric \ce{pCO2} and thus climate (Figure 3). 

\begin{figure*}
\begin{center}
\includegraphics[width=5cm]{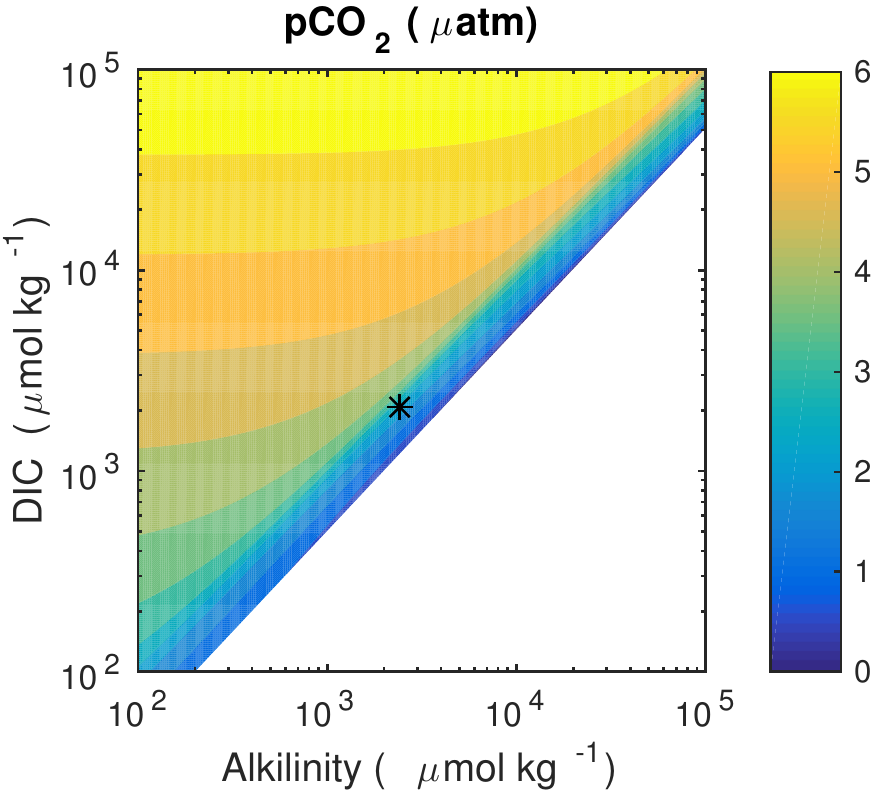} \includegraphics[width=5cm]{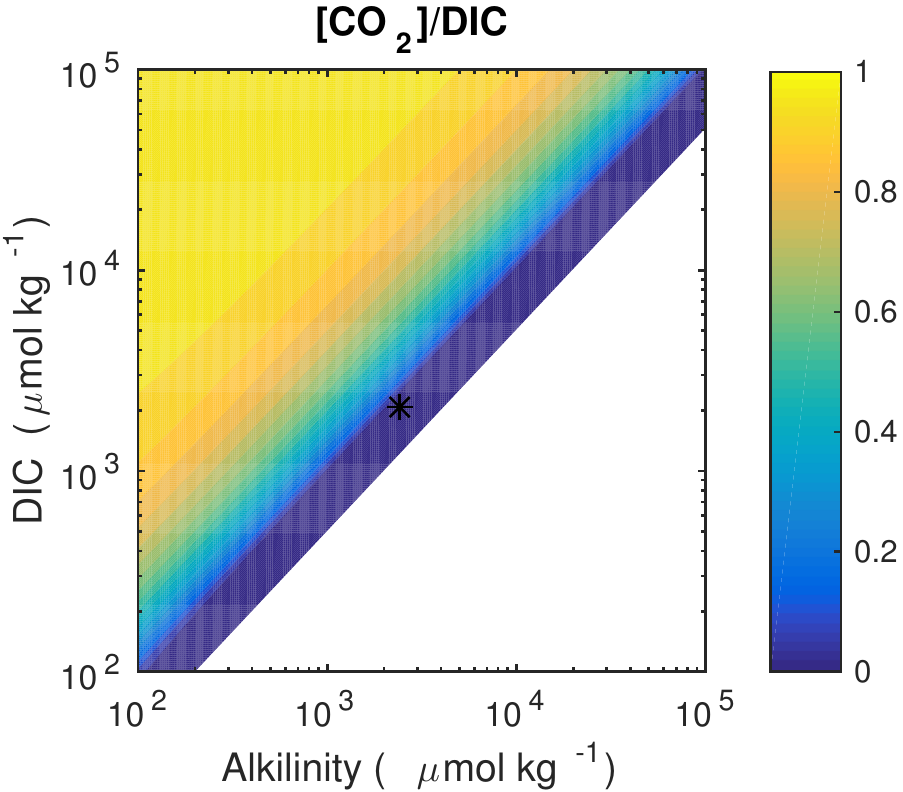}

\includegraphics[width=5cm]{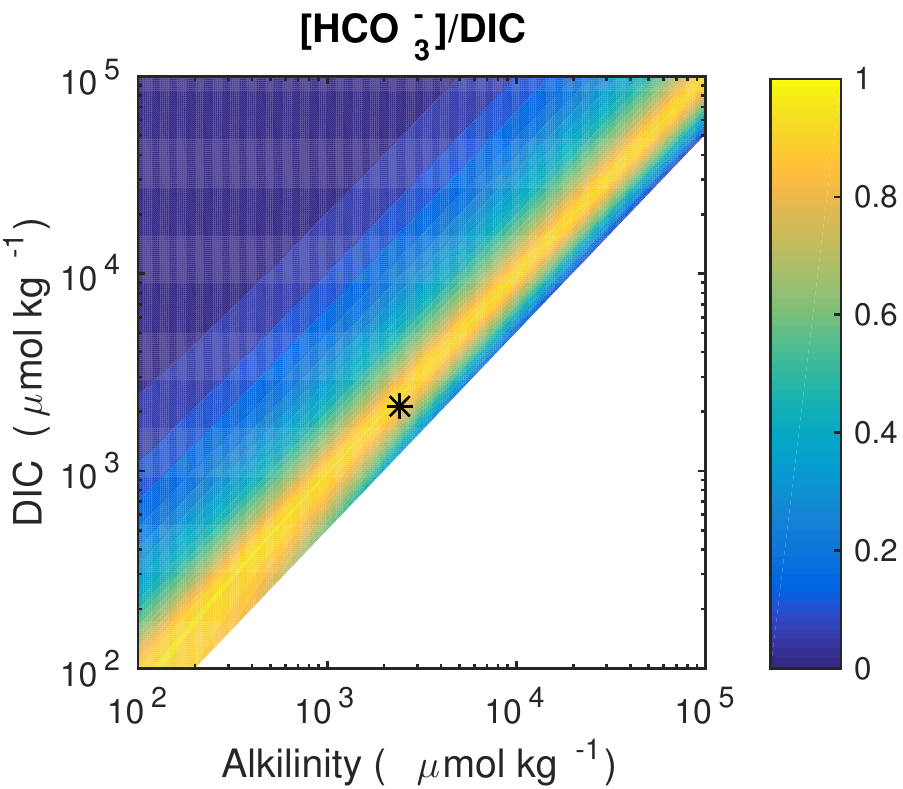} \includegraphics[width=5cm]{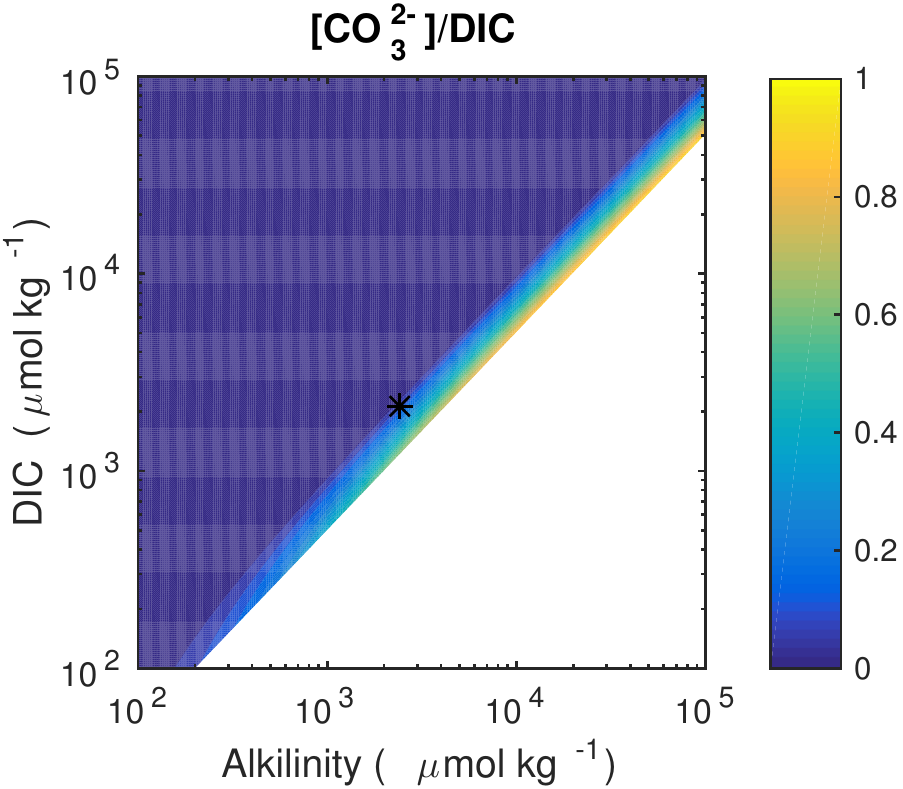}
\end{center}
\caption{Atmospheric carbon dioxide and dissolved species concentrations as a function of DIC and Alk. There is a tendency to think of the total amount of carbon in the atmosphere-ocean determining the strength of the \ce{CO2} greenhouse, but quite clearly alkalinity exerts equal leverage. Figures made with the Zeebe code\cite{ZeebeWolfGladrow}}
\end{figure*}

Plant roots increase soil \ce{pCO2} and secrete organic acids which will dissolve rock. 

In the ocean, high alkalinity means that \ce{[CO3^{2-}]} will increase to the point where precipitation of \ce{CaCO3} is thermodynamically favoured, and precipitation removes both DIC and alkalinity. Thus, the consequence of an alkalinity flux from weathering is removal of inorganic carbon from the atmosphere-ocean system and deposition in rock. The saturation product  \ce{\Omega = ([Ca^+][CO3^{2-}])_{sat}} describes when precipitation is thermodynamically favoured, but this is kinetically inhibited. Authigenic (abiological) precipitation requires $\Omega = 30$, but on Earth biology rules the roost yet again as organisms that build calcium carbonate shells are able to do so at $\Omega = 3$. Just as biology enhances weathering, it enhances carbonate deposition, and living Earth has lower \ce{pCO2} than her sterile equivalent would. 

Carbonate rocks will ultimate be destroyed through metamorphism, either of during subduction of ocean crust or in regional metamorphism of carbonate rocks uplifted to the continents. This recycled carbon contributes a larger part of the oft-referred to ``volcanic carbon'' source than juvenile carbon. Geological processes are thus involved in the partitioning between atmosphere-ocean and solid planet reservoirs. 

Now, the transition to Venus: first, the deposition of carbonates is aqueous, so with the evaporation and loss of the oceans none can be deposited; second, at high temperatures, calcium carbonate will break down \ce{CaCO3 -> CaO + CO2}. This reaction is operated industrially on Earth in lime kilns. The required temperature of around 1200K is quite reasonable for the surface of Venus in a runaway greenhouse (the hot dry state) before water is lost to space. 

In summary, to understand the first order controls of atmospheric \ce{CO2}, one must understand planetary climate, ocean chemistry, surface weathering processes and the action of life. 

\subhead{Di-nitrogen}

The geological cycle of nitrogen, hence the possible variation of atmospheric di-nitrogen inventory, have long been under-appreciated. However, modern whole-Earth nitrogen budgets suggest that Earth's mantle contains more nitrogen than the atmosphere and that mantle nitrogen is of subduction origin. An atmospheric mass worth of nitrogen can be subducted in around a billion years, so variations in the atmospheric budget have to be considered \cite{Goldblatt2013, Johnson2015}. 

This mantle nitrogen is biological in origin. The only way that large amounts of di-nitrogen can be fixed is through biological nitrogen fixation, to make ammonium (\ce{NH4^+}). Ammonium  has a similar ionic radius to potassium ion, so will readily substitute into K-rich minerals (particularly clays). Stable in a geological setting, it may then be subducted. Noble gas isotopes indicate that the mantle nitrogen (one to a few times the amount in the atmosphere) is indeed of subducted origin, and not primordial. 

As with carbon dioxide, there is a prima facia case that a mix of biological, geochemical and geological processes control the atmospheric inventory. 

\subhead{Water}

Water may be somewhat unique amongst the atmospheric gases in that its controls do appear to be physical, dominated by evaporation-condensation: warmer temperatures leads directly to higher partial pressure. Feedback processes in interaction with the radiation field give rise to the number of climate states discussed above. 

There are other considerations though. Existence of an ocean requires: initial delivery of water to a planet inside the snowline; failure to lose that ocean via hydrogen escape; failure to subduct all water to the mantle through hydration and subduction of rocks (which does happen to an extent on Earth).



\section*{Conclusion}

Were Earth to have only water in its atmosphere, it would today exist perilously close to both the inner and outer bounds of the conventionally defined habitable zone. In dynamical systems terms, there are several overlapping stable steady states of climate near present solar constant. The state with liquid water and modest temperature is narrow.

Expanding the habitable zone requires other gases in the atmosphere: conceptually, a greenhouse gas when there is little incoming sunlight and a background gas to either scatter sunlight when that is abundant, or broaden the absorption of greenhouse gasses when sunlight is scarce. The boundaries of the climate states are determined by atmospheric physics with abundances of these gases as a free parameter. On Earth, appeals to physical feedbacks to control carbon dioxide or di-nitrogen inventories fail: those free parameters are controlled by geology, geochemistry and life itself. 

Life, we see, has entered the business of controlling the boundaries of its own habitat in space and time, niche construction in ecological parlance. As we view Gaia through the atmosphere, we see that habitability and inhabitance are inseparable.

\subhead{Acknowledgements} This work was funded by an NSERC discovery grant. Thanks to Ben Johnson for comments on the manuscript.  

\bibliographystyle{unsrt}

\end{abstracttext}

\end{document}